\documentclass{article}

\usepackage{spconf,amsmath,graphicx}
\usepackage[utf8]{inputenc}
\usepackage[english]{babel}
\usepackage{xcolor}
\usepackage{booktabs}
\usepackage{siunitx}
\usepackage{lineno}
\usepackage{float}

\title{Audio-visual fine-tuning of audio-only ASR models}
\name{Avner May$^1$, Dmitriy Serdyuk$^1$, Ankit Parag Shah$^2$\sthanks{Worked on project as intern at Google, summer 2022.}, Otavio Braga$^1$, Olivier Siohan$^1$}
\address{$^1$Google, Inc. \;\;\;\;\;\;\; $^2$Carnegie Mellon Compute Science Department}

\begin{document}
%\ninept
%
\maketitle

\begin{abstract}
Audio-visual automatic speech recognition (AV-ASR) models are very effective at reducing word error rates on noisy speech, but require large amounts of transcribed AV training data.
Recently, audio-visual self-supervised learning (SSL) approaches have been developed to reduce this dependence on transcribed AV data, but these methods are quite complex and computationally expensive.
In this work, we propose replacing these expensive AV-SSL methods with a simple and fast \textit{audio-only} SSL method, and then performing AV supervised fine-tuning.
We show that this approach is competitive with state-of-the-art (SOTA) AV-SSL methods on the LRS3-TED benchmark task (within 0.5\% absolute WER), while being dramatically simpler and more efficient (12-30x faster to pre-train).
Furthermore, we show we can extend this approach to convert a SOTA audio-only ASR model into an AV model.
By doing so, we match SOTA AV-SSL results, even though no AV data was used during pre-training.
\end{abstract}


\section{Introduction}
\label{sec:intro}

Audio-visual automatic speech recognition (AV-ASR) models~\cite{Matthews2002-iv,Assael2016-zg,makino19} leverage the video of a speaker's moving lips, in addition to the audio, to improve recognition performance.
These AV-ASR models are typically able to reduce the word error rate (WER) on noisy speech by 2-3x~\cite{robust_avhubert}, relative to audio-only models, but require large amounts of transcribed AV data to train~\cite{makino19}.
To reduce the dependence of these AV-ASR models on transcribed AV training data, a variety of AV self-supervised learning (SSL) methods (e.g., \cite{robust_avhubert,avhubert,uhubert,av_data2vec,vatlm}) have been proposed, which make use of unsupervised AV data for pre-training.
However, these methods are generally quite complex and computationally expensive.
For example, one state-of-the-art method, u-HuBERT~\cite{uhubert}, requires multiple iterations of pre-training, and can take up to two weeks to pre-train.

\begin{figure}[t]
\includegraphics[width=0.48\textwidth]{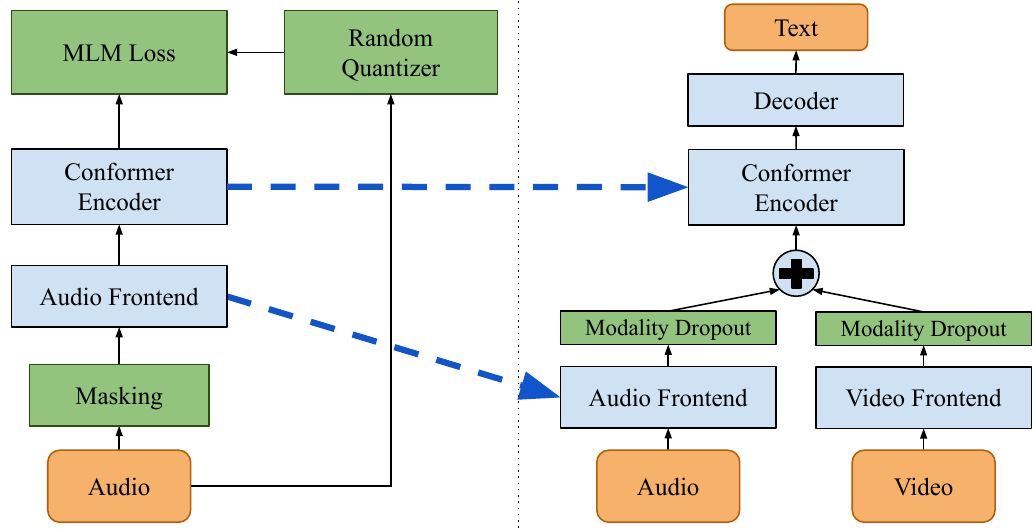}
\caption{
\textit{A high-level summary of our proposed FAVA method.}
The first stage of training (left) pre-trains the audio front-end and the encoder using the BEST-RQ~\cite{bestrq} pre-training method.
The second stage (right) performs audio-visual supervised learning, where the parameters for the audio front-end and encoder are initialized from the stage 1 model.
}
\label{fig:fava}
\end{figure}

In this work, we ask what performance is achievable if we replace these complex and slow AV-SSL methods with a much simpler and faster \textit{audio-only} SSL method.
We call this approach FAVA---\underline{F}ine-tuning \underline{A}udio-\underline{V}isual models from \underline{A}udio-only models.
This method has two stages, as illustrated in Figure~\ref{fig:fava}:
First, we use the BEST-RQ audio-only SSL method \cite{bestrq} for pre-training.
We then use this pre-trained model to initialize an AV-ASR model, which we fine-tune using a small transcribed AV dataset.
We use an early-fusion architecture for the AV-ASR model, with convolutional audio and video front-ends, a Conformer~\cite{conformer} encoder, and an RNN-T decoder~\cite{rnnt}.

We evaluate FAVA by comparing it empirically with state-of-the-art AV-SSL methods in the literature~\cite{robust_avhubert,uhubert,av_data2vec,vatlm}, and with a thorough set of baselines we implemented.
For fair comparison with these AV-SSL methods, we follow the same experimental protocol of using VoxCeleb2~\cite{voxceleb2} as the unsupervised dataset, LRS3-TED~\cite{ted} as the supervised dataset, and evaluating on the LRS3-TED test set.
To better understand the performance of FAVA, we also implement a number of strong audio-only and AV supervised and self-supervised baselines.

Our experimental results demonstrate that FAVA approaches the performance of state-of-the-art AV-SSL methods, while being dramatically faster; it also meaningfully outperforms the strong audio-only and AV baselines we implemented.
Relative to state-of-the-art AV-SSL methods, FAVA attains a WER within 0.5\% (absolute) on clean audio, and is 12-30x faster to pre-train.
Relative to the baselines we implemented, we show FAVA attains 5\% and 57\% relative reduction in WER compared to audio-only self-supervised models on clean and noisy audio, respectively.
It also attains 41\% and 22\% relative reduction in WER compared to AV trained-from-scratch models on clean and noisy audio.

The FAVA method can easily be extended to convert any state-of-the-art audio-only ASR model into an audio-visual model, which attains better performance on noisy audio while maintaining or improving the model's strong performance on clean audio.
This allows audio-visual models to directly benefit from the huge datasets and model sizes now being used to train audio-only ASR models~\cite{usm,whisper,meta_mms}, and from any improvements in the training of these models.
To demonstrate the potential of this approach, we show that we can convert a version of the Google Universal Speech Model (USM)~\cite{usm} into an AV model, which matches the performance of the state-of-the-art AV-SSL methods without using any audio-visual data during pre-training.

The rest of the paper is organized as follows:
In Section~\ref{sec:related_work} we review related work and background.
In Section~\ref{sec:FAVA} we present our method, FAVA.
In Sections~\ref{sec:experimental_setup} and \ref{sec:results} we discuss our experimental setup and results, respectively.
In Section~\ref{sec:extensions} we show how we can extend FAVA, and easily convert state-of-the-art audio-only models into AV models.
We conclude in Section~\ref{sec:conclusion}.

\section{Related Work and Background}
\label{sec:related_work}
In this section we summarize work on self-supervised learning (SSL) for both audio-only (Section~\ref{sec:related_work_audio}) and audio-visual (Section~\ref{sec:related_work_av}) speech recognition, and discuss how our work builds on this work by leveraging a simple and fast audio-only SSL method~\cite{bestrq} to pre-train AV-ASR models.
% The goal of these methods is to leverage large unsupervised datasets to improve the performance of speech recognition models that are trained on limited amounts of transcribed data.

\subsection{Audio-only self-supervised learning}
\label{sec:related_work_audio}
Most audio-only self-supervised learning methods~\cite{bestrq, wav2vec2,w2vbert,hubert,data2vec} mask random segments of an input utterance, and train a model to predict targets corresponding to the frames within those masked segments.
The targets are generally based on the intermediate activations, at the masked frames, of a model which received the unmasked utterance as input.
The model used to generate these targets can be the same as the model being trained~\cite{wav2vec2,w2vbert}, or a model whose parameters lag behind the model being trained (e.g., from a prior checkpoint~\cite{hubert}, or from an exponential moving average of the model's parameters~\cite{data2vec}).
The loss function used could be a masked language modeling (MLM) cross-entropy loss~\cite{bestrq,w2vbert,hubert}, a regression loss~\cite{data2vec}, or a contrastive loss~\cite{wav2vec2,w2vbert}.

There are a number of issues with using intermediate activations of an evolving model for determining the targets during pre-training, including being complex and difficult to train, as well as being computationally expensive.
These approaches are susceptible to failure modes, known as ``model collapse''~\cite{data2vec}, where the model simply learns to predict constant values, thus making the self-supervised learning task trivial.
This makes the optimization trickier and more sensitive to hyperparameters, and also makes the pre-training harder to reason about.
To mitigate this issue, one can do several iterations of pre-training, where for each iteration the targets are computed using a \textit{fixed} checkpoint~\cite{hubert};
but in this case, training can become quite expensive, as several iterations of pre-training are often needed to get the best performance.
These iterative methods are also expensive and error prone from an engineering perspective, as generally an ML engineer would need to monitor the training process, and manually start each iteration using a checkpoint from the previous iteration.

To address the above issues with using targets based on intermediate activations, the BEST-RQ method~\cite{bestrq} uses pre-training targets that are determined by a random quantization of the acoustic features for a masked frame.
It then uses a simple MLM cross-entropy loss function with these targets as the self-supervised learning objective.
Because the pre-training targets are a fixed function of the acoustic features, the method is robust to model collapse.
BEST-RQ is simple and efficient, and attains state-of-the-art performance on the LibriSpeech~\cite{librispeech} benchmark task, when pre-training on the LibriLight dataset~\cite{librilight}.
It is used as the pre-training method for the recent state-of-the-art audio-only Google USM model~\cite{usm}.
Our proposed method, FAVA, leverages the speed and simplicity of BEST-RQ for pre-training AV-ASR models.

\subsection{Audio-visual self-supervised learning}
\label{sec:related_work_av}
Audio-visual self-supervised learning methods are often quite similar to audio-only ones, but take both audio and video as input.
For example, AV-HuBERT~\cite{avhubert,robust_avhubert} and u-HuBERT~\cite{uhubert} adapt the audio-only method HuBERT~\cite{hubert} to the audio-visual setting, while AV-data2vec is based on the data2vec method~\cite{data2vec}.
As mentioned above, these methods can be quite complex and computationally expensive.

One recent work, \textsc{Vat}LM~\cite{vatlm}, pre-trains a model using a combination of audio-visual, audio-only, audio-text, and text-only data.
The targets during pre-training are derived from an AV-HuBERT model~\cite{avhubert}, and thus \textsc{Vat}LM depends on access to a strong pre-trained model.

One paper which shares our goal of leveraging audio-only self-supervised models for training AV-ASR models is~\cite{pan2022leveraging}.
This method uses a late-fusion architecture to combine the outputs of an audio-only model trained using wav2vec 2.0~\cite{wav2vec2} with those of a visual front-end trained using a modification of Moco v2~\cite{moco_v2} to train an AV-ASR model.
However, the training process in~\cite{pan2022leveraging} is complex and expensive, requiring multiple stages of modality-specific pre-training and supervised training, followed by multi-modal supervised training.
Our approach, in contrast, uses a more parameter-efficient early-fusion architecture, and only requires a single round of audio-only pre-training followed by audio-visual fine-tuning.

\section{FAVA Method}
\label{sec:FAVA}
We now give an overview of the stages of the FAVA method (Section~\ref{sec:fava_stages}), and of the components of the model architectures we use for this method (Section~\ref{sec:architecture}).

\subsection{Training stages}
\label{sec:fava_stages}
The FAVA method has two stages, as illustrated in Figure~\ref{fig:fava}: audio-only self-supervised learning, followed by audio-visual supervised learning.

\vspace{0.1in} \noindent \textbf{Audio-only self-supervised learning}: 
In the first stage, we perform audio-only self-supervised learning using the BEST-RQ method~\cite{bestrq}.
We picked BEST-RQ because it is a simple, fast, and robust method, as discussed in Section~\ref{sec:related_work_audio}.
% Using BEST-RQ, we are able to pre-train a model in 12 hours on a 16x16 TPU cluster.
% This is in contrast to AV-HuBERT~\cite{avhubert}, which has a total training time of approximately 2 weeks on 32-64 V100 GPUs, using the same pre-training data.

We use the BEST-RQ hyperparameters from the original paper~\cite{bestrq};
we use a target codebook of size 8192, and a masked segment length of 400 ms (40 frames, at a frame rate of 100 Hz), where 1\% of frames are randomly selected as the beginning of a masked segment.
To reduce the variance between the runs, we use the same parameters (random projection and codebook matrices) for the BEST-RQ random quantization for all of our experiments.
We generate pre-training targets at a frame rate of 25 Hz by stacking four consecutive acoustic feature vectors (frame rate 100 Hz), which we then randomly quantize.

\vspace{0.1in}  \noindent \textbf{Audio-visual supervised learning}:
Next, we perform audio-visual supervised learning with the RNN-T loss~\cite{rnnt,zhang_2020}, initializing the audio front-end and the encoder from the previous stage's model, and initializing the visual front-end and decoder randomly.
In our experiments, we use an early-fusion AV architecture, as illustrated in Figure~\ref{fig:fava}.

To prevent the model from relying too much on any single modality, we use modality dropout, as in the AV-HuBERT and u-HuBERT methods~\cite{robust_avhubert,avhubert,uhubert};
with probability 0.25 we zero out the audio features (output of audio front-end) for the entire utterance, with probability 0.25 we zero out the video features (output of video front-end), and with probability 0.5 we zero out neither.

To make our AV models more robust, during fine-tuning we utilize SpecAugment~\cite{specaug}, and also add different types of noise (e.g., background cafe noise, music) to the input utterances as a form of multi-style training~\cite{mtr}.

\subsection{Model architecture}
\label{sec:architecture}
We now describe in detail the various components of the model architectures we use in the FAVA method, including the audio and video front-ends, the encoder, and the decoder.

\vspace{0.1in} \noindent \textbf{Audio front-end}:
The inputs to the audio front-end are 80-dimensional log-mel filterbank feature vectors, at a frame-rate of 100~Hz (25~ms window size, 10~ms stride).
The audio front-end consists of 2 convolution layers, with `(time, frequency, in channels, out channels)' filter shapes of size $(3, 3, 1, 128)$ and $(3, 3, 128, 32)$, and `(time, frequency)' strides of size $(2, 2)$ and $(2, 2)$, for the two layers (respectively).
Thus, the output of the audio front-end has a framerate of 25~Hz, and a dimension of 20x32 (20 along frequency dimension, with 32 output channels), which we flatten and then linearly project down to 512 output dimensions (to allow summing to the output of the video front-end).
% task.encoder.input_converters[1].sub.filter_shapes : [(3, 3, 1, 128), (3, 3, 128, 32)]
% task.encoder.input_converters[1].sub.filter_strides : [(2, 2), (2, 2)]

\vspace{0.1in} \noindent \textbf{Video front-end}:
The inputs to the video front-end are the video of the speaker, cropped around their lips\footnote{We use FaceNet~\cite{facenet} for detecting the region around the lips, as in~\cite{makino19}.}, at a resolution of 128x128 (with 3 RGB color channels), sampled at a frame rate of 25~Hz.
We normalize the video tensor to have values in $[-1, 1]$.
We use a convolutional video front-end for all our experiments, following the `(2+1)D ConvNet' architecture from~\cite{vit}.
This is a VGG-style architecture that alternates (5 times) between spatial and temporal convolutions of sizes $(1, 3, 3)$ and $(3, 1, 1)$, respectively (in `(time, height, width)' format).
The output is 512 dimensional, and remains at a frame rate of 25 Hz.
% task.encoder.extractors[0].video_frontend.kernel_sizes : [[1, 3, 3], [3, 1, 1], [1, 3, 3], [3, 1, 1], [1, 3, 3], [3, 1, 1], [1, 3, 3], [3, 1, 1], [1, 3, 3], [1, 1, 1]]
% task.encoder.extractors[0].video_frontend.out_channels : [23, 64, 230, 128, 460, 256, 921, 512, 460, 512]

\vspace{0.1in} \noindent \textbf{Encoder}:
We use a 17-layer Conformer~\cite{conformer} encoder with 8 attention heads, 512 hidden units (input/output of self-attention layer), and 2048 hidden units for the feed-forward network.
% task.encoder.extractors[10].sub.trans_atten_tpl.num_heads : 8
% task.encoder.extractors[13].sub.fflayer_start_tpl.input_dim : 512
% task.encoder.extractors[11].sub.fflayer_end_tpl.input_dim : 512
% task.encoder.extractors[11].sub.input_dim : 512
% task.encoder.extractors[14].sub.fflayer_end_tpl.hidden_dim : 2048

\vspace{0.1in} \noindent  \textbf{Decoder}:
We use an RNN-T decoder~\cite{rnnt,zhang_2020}, with a two-layer LSTM (1280 hidden units, 128 embedding dim) as the predictor network, and a single-hidden layer feed-forward network (640 hidden units) as the joiner network.
The output of the decoder are logits corresponding to 4096 WordPieces~\cite{wordpieces}.
% task.decoder.joint_network_size : 640
% task.decoder.rnn_cell_dim : 1280
% task.decoder.num_classes : 4096
% task.decoder.emb_dim : 128
\\
\newline
Our audio-only and AV models have a total 128M and 135M parameters, respectively.
% With the above components combined, 
% Summing the parameter counts from the above components, our audio-only and AV models have a total 128M and 135M parameters, respectively.
% , accounting for parameters from all the above-mentioned components.

\section{Experimental Design}
\label{sec:experimental_setup}
In this section, we describe the protocols we use for our experiments comparing FAVA with published results (Section~\ref{sec:exp_setup_published}), as well as with a thorough set of baselines we implemented (Section~\ref{sec:exp_setup_baselines}).

\subsection{Comparison to published AV-SSL results}
\label{sec:exp_setup_published}
To allow for a fair comparison between FAVA and published AV-SSL results, we make sure to use the same datasets for training and evaluation.
More specifically, we use the VoxCeleb2~\cite{voxceleb2} dataset as the unsupervised dataset, the LRS3-TED~\cite{ted} pre-train and train-val datasets (combined) as the supervised dataset, and the LRS3-TED test set as the primary evaluation set.
We now describe these datasets in more detail:

\begin{itemize}
    \item \textbf{VoxCeleb2~\cite{voxceleb2}} contains about 2400 hours of untranscribed speech from YouTube videos of celebrities.
    We extract a $\sim\!1300$ hour subset of English language videos which we use for self-supervised pre-training, as in~\cite{robust_avhubert,avhubert,uhubert,av_data2vec,vatlm}.
    \item \textbf{LRS3-TED~\cite{ted} (pre-train set)} contains about 400 hours of transcribed English TED talks.
    We segment the utterances that are longer than 15 seconds into segments with a maximum length of 15 seconds (splitting at word boundaries, following the protocol used by AV-HuBERT~\cite{avhubert}).
    We use this data for both self-supervised and supervised training.
    \item \textbf{LRS3-TED~\cite{ted} (train-val set)} contains about 30 hours of transcribed English TED talks.
    We randomly extract around 1200 utterances (about one hour) to serve as a held-out set during training, following the protocol used by AV-HuBERT~\cite{avhubert}.
    We use this data for both self-supervised and supervised training.
    \item \textbf{LRS3-TED~\cite{ted} (test set)} contains about 1 hour of transcribed English TED talks.
    We use this set to evaluate our models.
    Note that we also create a noisy version of this test set by adding 0~dB babble noise sampled from the NoiseX~\cite{noisex} dataset.
    This noisy test set allows us to measure the WER improvements AV models are able to attain on noisy audio relative to audio-only ASR models.
    \item \textbf{YT-DEV-18~\cite{serdyuk21} (development set)} is a professionally transcribed dataset of 32 hours of YouTube videos.
    Similarly to the LRS3-TED test set, we create a noisy version of this development set by adding 0~dB babble noise.
    We use these development sets for checkpoint selection during AV supervised fine-tuning.
    In particular, we select the checkpoint which attains the best average WER across the clean and noisy versions of this dataset.
\end{itemize}

Note that both u-HuBERT~\cite{uhubert} and \textsc{Vat}LM~\cite{vatlm} also leverage the TED-LIUM 3~\cite{tedlium3} dataset, which contains 452 hours of audio, for additional data during pre-training.
\textsc{Vat}LM additionally uses 3846 hours of unlabeled English YouTube audio data from the XL subset of GigaSpeech~\cite{gigaspeech}, and 600 million English sentences (just text, no audio) from the Cantab-TEDLIUM Release 1.1 dataset~\cite{cantab}.

\begin{table*}[t]
    \centering
    \caption{\textit{Comparison between FAVA and results in the literature, as well as baselines we implemented.}
    We report WERs on clean and noisy versions of the LRS3-TED~\cite{ted} test set.
    We additionally report the number of hours of training data (audio-only/AV, supervised/unsupervised) used to train each of these models.
    Below, `TFS' stands for `train-from-scratch', `PT' stands for `pre-train', and `FT' stands for `fine-tune'.
    Note ($\dagger$) that u-HuBERT~\cite{uhubert} and robust AV-HuBERT~\cite{robust_avhubert} use different noise sources than we do to generate the noisy version of the test set, making their noisy WERs not strictly comparable to ours.
    }
    \begin{tabular}{lrrrrS[table-format=2.2]S[table-format=2.2]}
        \toprule
         & \multicolumn{2}{c}{Unsupervised (hrs)} & \multicolumn{2}{c}{Supervised (hrs)} & \multicolumn{2}{c}{WER (\%)}  \\
        \bfseries Model     & { A} & { AV} & {A} & {AV} & {Clean} & {Noisy} \\
        \midrule
        \multicolumn{7}{l}{\hspace{3pt} \emph{Published Audio-Visual SSL Results}} \\
        Robust AV-HuBERT~\cite{robust_avhubert} & - & 1759 & - & 433 & 1.4 & 5.8 $\dagger$ \\
        u-HuBERT~\cite{uhubert}       & 452 & 1759 & - & 433 & 1.3 & 4.6 $\dagger$ \\
        AV-data2vec~\cite{av_data2vec}    & - & 1759 & - & 433 & 1.3 & \textemdash \\
        \textsc{Vat}LM~\cite{vatlm} & 4298 & 1759 & - & 433 & 1.2 & \textemdash \\
        \midrule
        \hspace{3pt} \emph{Published TFS Baselines} \\
        u-HuBERT~\cite{uhubert} (audio-only)      & - & - & 433 & - & 4.0 & 37.3 $\dagger$ \\
        u-HuBERT~\cite{uhubert} (AV)       & - & - & - & 433 & 3.6 & 15.9 $\dagger$ \\
        \midrule
        \hspace{3pt} \emph{Our Baselines} \\
        Audio-only TFS   & - & -  & 433 & - & 3.6 & 15.3 \\
        Audio-only PT+FT & 1759 & - & 433 & - & 2.0 & 15.3 \\
        AV TFS           & - & - & - & 433 & 3.3 & 8.5 \\
        AV PT+FT         & - & 1759 & - & 433 & 2.1 & 6.8 \\
        \midrule
        \hspace{3pt} \emph{Our Results} \\
        FAVA           & 1759 & - & - & 433 & 1.7 & 6.6 \\
        \midrule
        \multicolumn{7}{l}{\hspace{3pt} \emph{Our Results: FAVA Extensions}} \\
        % FAVA + FT      & 1759 & - & 433 & 433 & 1.8 & 6.8 \\
        USM~\cite{usm}   & 12M & - & 5000 & - & 2.5 & 9.0 \\
        USM~\cite{usm} + audio-only FT   & 12M & - & 5433 & - & 1.3 & 9.2 \\
        USM~\cite{usm}  + AV FT      & 12M & - & 5000 & 433 & 1.3 & 6.2 \\
        % Audio-only USM   & 12M & - & 5000 & - & 1.2 & 7.9 \\
        \bottomrule
    \end{tabular}
    \label{tab:av_asr}
\end{table*}

\begin{table}[t]
    \centering
    \caption{\textit{Comparison of computation cost for SSL methods.}
    We report the number of days of pre-training for published AV-SSL baselines and FAVA, and the type of hardware used.}
    \begin{tabular}{lrr}
        \toprule
        Method & \# days & Hardware used \\
        \midrule
        \hspace{3pt} \emph{Published Baselines} \\
        *-HuBERT~\cite{avhubert,robust_avhubert,uhubert} & 10 - 15  & 32-64 V100 (32GB) \\
        % 2/3 days for first 4 iterations, 2.4/3.6 for large
        AV-data2vec~\cite{av_data2vec} & 10 - 12 & 64 V100 (32GB)\\
        \textsc{Vat}LM~\cite{vatlm} & 6 & 32 A100 (80GB) \\
        % \midrule
        % \hspace{3pt} \emph{Our Baselines} \\
        % Audio-only PT+FT & $\sim 0.5$ & 256 v3 TPU (32GB)  \\
        % AV PT+FT & $\sim 1.5$ & 256 v3 TPU (32GB)  \\
        \midrule
        \hspace{3pt} \emph{Our Results} \\
        FAVA & 0.5 & 256 v3 TPU (32GB)  \\
        \bottomrule
    \end{tabular}
    \label{tab:training_time}
    % \vspace{-0.08in}
\end{table}

\subsection{Comparison to our own baselines}
\label{sec:exp_setup_baselines}
To better understand the performance of FAVA, we implement a number of natural baselines:
audio-only and audio-visual models trained from scratch on the LRS3-TED supervised set,
and audio-only and AV models pre-trained on VoxCeleb2 and LRS3-TED, and fine-tuned on the LRS3-TED supervised set.

In terms of the model architectures, the audio-only and audio-visual models use the same model architecture described in Section~\ref{sec:architecture}, with the only difference being that the audio-only architecture does not include a video front-end or modality dropout.

In terms of the pre-training algorithms, we perform BEST-RQ~\cite{bestrq} pre-training for the audio-only models, and an audio-visual version of BEST-RQ for the audio-visual models.
For the AV version of BEST-RQ, we use the same loss and targets as the audio-only version, but also allow the model to utilize the video input during pre-training.
We use modality dropout and video-masking during pre-training, where we mask the video---during the same time segments we mask the audio---by replacing each masked video frame with a random frame from the same video.
These models all use SpecAugment~\cite{specaug} and artificially add noise to the input utterances (as described in Section~\ref{sec:fava_stages}) during the supervised portion of the training process.

\section{Results}
\label{sec:results}
Below, we present results comparing FAVA to state-of-the-art AV-SSL methods in the literature (Section~\ref{sec:results_published}), and to the set of thorough baselines we implemented (Section~\ref{sec:results_baselines}).
In comparison to published AV-SSL methods, we show that FAVA attains performance within 0.5\% WER (absolute) of state-of-the-art audio-visual self-supervised learning methods, while being 12-30x faster to pre-train.
Additionally, we show FAVA meaningfully outperforms the strong audio-only and AV baselines we implemented.
% Compared to our audio-only self-supervised learning baselines, FAVA performs slightly better on clean audio (5\% relative reduction in WER), while reducing the WER dramatically on noisy audio (57\% relative reduction in WER).
% Lastly, compared to our audio-visual supervised baselines (trained from scratch), FAVA attains considerable reductions in WER on both clean and noisy test sets (41\% and 22\% relative reduction in WER, respectively).
% These results demonstrate that FAVA is a fast and simple method for attaining strong AV-ASR results.

\subsection{Comparison to published AV-SSL results}
\label{sec:results_published}
We compare FAVA---in terms of both recognition performance (top of Table~\ref{tab:av_asr}), and pre-training time (Table~\ref{tab:training_time})---to several state-of-the-art AV-SSL methods.
% 

% All of these AV-SSL methods pre-train on VoxCeleb2 and TED, and fine-tune on TED.

% In the top part of Table~\ref{tab:av_asr}, we show the performance of three state-of-the-art audio-visual self-supervised learning methods, which all pre-train on VoxCeleb2 and TED, and fine-tune on TED.

In terms of recognition performance (Table~\ref{tab:av_asr}), FAVA attains a WER of 1.7\% and 6.6\% on the clean and noisy LRS3-TED test sets (respectively), compared to 1.2\%-1.3\% and 4.6\% for these AV-SSL methods.
While these AV-SSL methods attain approximately 30\% lower WER than FAVA (on both clean and noisy test sets), it is difficult to pinpoint the exact source of these performance improvements.
One potential source of discrepancy is that 12\% of LRS3-TED test utterances have been deleted from YouTube since the dataset was created in 2018, and we were unable to evaluate on these utterances for legal reasons.
Furthermore, our noisy LRS3-TED test set is constructed using 0~dB babble noise from the NoiseX~\cite{noisex} dataset, whereas AV-HuBERT~\cite{robust_avhubert} and u-HuBERT~\cite{uhubert} create babble noise by mixing 30 random utterances from LRS3-TED~\cite{ted}.

In terms of pre-training time, we show in Table~\ref{tab:training_time} that FAVA is 12-30x faster than published AV-SSL methods.
FAVA pre-training takes 12 hours, which is faster than the baselines (6-15 days) even taking into account the higher number of accelerators we use.
The fact that we don't use video during pre-training allows us to fit a larger mini-batch per chip in terms of seconds of audio (256 vs. 40), and to perform faster model updates.

% \vspace{-0.08in}
\subsection{Comparison to our own baselines}
\label{sec:results_baselines}
Due to the difficulty of performing apples-to-apples comparisons with the published results, we implemented a set of strong baselines corresponding to self-supervised and supervised models, for both audio-only and audio-visual input modalities.
Surprisingly, we see in Table~\ref{tab:av_asr} that even though FAVA doesn't see the visual modality during pre-training, it performs better than all of these baseline methods, attaining a WER of 1.7\% and 6.6\% on the clean and noisy LRS3-TED test sets (respectively), compared to at best 2.0\% and 6.8\% across the baseline methods.

More specifically, compared to our audio-only self-supervised learning baselines, FAVA performs slightly better on clean audio (5\% relative reduction in WER), while reducing the WER dramatically on noisy audio (57\% relative reduction in WER).
Additionally, compared to our audio-visual supervised baselines (trained from scratch), FAVA attains considerable reductions in WER on both clean and noisy test sets (41\% and 22\% relative reduction in WER, respectively).

We note that our baselines are strong:
our trained-from-scratch baselines perform better across the board than the trained-from-scratch results reported in~\cite{uhubert} (summarized in Table~\ref{tab:av_asr}), and our pre-trained + fine-tuned results attain large gains (generally 20-45\% relative reduction in WER) compared to these strong trained-from-scratch baselines.
Furthermore, we see that the audio-visual models (TFS and PT+FT) perform similarly to the audio-only models on the clean test set, while performing significantly better on the noisy test set, as expected.

% Compared to these strong baselines, we (surprisingly) see that FAVA is able to attain better performance on both the clean and noisy tests, even though it does not use the visual modality during pre-training.
% We hypothesize that FAVA (audio pre-training + AV fine-tuning) performs better than audio-visual pre-training and fine-tuning, because the pre-training task is too difficult to perform with just the video. Thus the video serves just as a distraction during pre-training.
% Interestingly, published AV-SSL results do not compare with this important baseline of ignoring the video during pre-training, raising the question of whether the video is actually helping during pre-training.

\section{FAVA extensions}
\label{sec:extensions}
We now demonstrate that a simple extension of the FAVA method allows us to convert state-of-the-art audio-only ASR models into AV models.
For this experiment, we take a 600 million parameter version of the Google USM model~\cite{usm}, and perform either audio-only or AV supervised fine-tuning using the LRS3-TED supervised training set.
This USM model was pre-trained on $12M$ hours of unsupervised multi-lingual YouTube videos, and fine-tuned on $5000$ hours of transcribed English YouTube videos.
Before the fine-tuning steps on the LRS3-TED training set, this audio-only model attains WERs of 2.5\% and 9.0\% on the clean and noisy LRS3-TED test sets, respectively.
% We take checkpoint ~140k from these experiments:
% https://tensorboard.corp.google.com/compare/audio_only:6666183103446905761,av:8787458466173310192/?enableScalarColumnCustomization=true#timeseries
As we show in Table~\ref{tab:av_asr}, audio-only and AV fine-tuning both result in WERs of 1.3\% on the clean test set, thus effectively matching the SOTA AV-SSL methods~\cite{robust_avhubert,uhubert,av_data2vec,vatlm}.
On the noisy test set, however, the audio-only fine-tuning attains a WER of 9.2\%, while the AV fine-tuning results in a WER of 6.2\%, a 31\% relative reduction in WER compared to the USM model without fine-tuning.
These results demonstrate the potential of using FAVA to convert audio-only ASR models into AV-ASR models, and thereby attain large improvements in the recognition performance on noisy audio.

% As we show in Table~\ref{tab:av_asr}, performing audio-only fine-tuning results in WERS of 1.3\% and 9.2\%, while performing AV fine-tuning results in WERS of 1.3\% and 6.2\% (on clean and noisy, respectively).
% Thus, both the audio-only and AV fine-tuned models effectively match the SOTA AV-SSL methods~\cite{robust_avhubert,uhubert,av_data2vec,vatlm} on clean audio.
% Furthermore, the 

% As we show in Table~\ref{tab:av_asr}, before AV fine-tuning
% which was pre-trained on $12M$ hours of unsupervised multi-lingual YouTube videos, and fine-tuned on $5,000$ hours of transcribed English YouTube videos.
% This model attains WERs of 2.5\% and 9.0\% on the clean and noisy LRS3-TED test sets, respectively.
% We then perform AV supervised fine-tuning of this model, which reduces the WERs to 1.2\% and 6.1\% (respectively), thus matching the state-of-the-art AV-SSL methods on clean audio.
% ---and further fine-tune it on the LRS3-TED training set (433 hours).
% As we show in Table~\ref{tab:av_asr}, this model achieves 1.2\% WER on clean data, thus matching the performance of the state-of-the-art AV-SSL methods.
% On noisy data, it attains 6.1\% WER (32\% relative reduction in WER compared to the audio-only USM model).
% These results demonstrate the potential of using FAVA to convert audio-only ASR models into AV-ASR models, and thereby improve their performance on noisy audio.

One limitation of this approach of adapting state-of-the-art audio-only models to AV, is that by fine-tuning the entire model we can change its behavior when it receives audio-only inputs.
This could be undesirable in cases where the audio-only model is known to perform very well across a range of settings, and regressions in performance would be problematic and difficult to fix.
One exciting direction for future work would be to use adapters (e.g., ~\cite{lora,thomas_2022}) when performing the AV supervised fine-tuning step, thereby keeping the audio-only performance of the model unchanged \textit{by design} when converting the model to AV.

\section{Conclusion}
\label{sec:conclusion}
In this work, we proposed FAVA, a simple method which performs audio-visual pre-training followed by audio-visual fine-tuning.
FAVA attains performance within 0.5\% WER (absolute) of state-of-the-art AV self-supervised learning method, even though it does not leverage any video data during pre-training.
We showed that we can easily extend FAVA to convert state-of-the-art audio-only ASR models into AV models.
With this approach, we were able to adapt a Google USM model~\cite{usm} to match the performance of the best AV-SSL methods.
This suggests that a promising direction going forward for AV-ASR research is to focus on adapting state-of-the-art audio-only models, instead of on developing ever-more complex AV pre-training methods.

% \section{Safe AI Principles}
% We are aware of the sensitive nature of the AV-ASR research
% and other AI technologies used in this work.
% Therefore, we ensure that this work abides by the Google AI Principles~\cite{noauthor_undated-lg}.

\bibliographystyle{IEEEbib}
\bibliography{refs}

\end{document}